\documentstyle[epsf]{article}
\newcommand{\bo}[1]{\mbox{\boldmath $ #1 $}}
\newcommand{\s}[1]{{\rlap/ #1}}

\begin{document}

\begin{center}
{\Large \bf
Renormalization prescriptions in effective theory \\
of pion-nucleon scattering
} \\

\vspace{4mm}

A.~Vereshagin \\
University of Bergen and St.-Petersburg State University \\
Alexander.Vereshagin@ift.uib.no\\
\end{center}

\begin{abstract}
Using the results presented in the talk
\cite{VVV-QFTHEP}
we illustrate the approach developed in
\cite{VVV}--\cite{AVVV2}
by the calculation of
$\pi N$
elastic scattering amplitude and explain how to derive corresponding
bootstrap conditions. In the end, we briefly mention some results of
numerical comparison with experimental data.
\end{abstract}

\section{Introduction}

The essence of our work is an attempt to develop a self-consistent
Dyson perturbation technique for the infinite component effective
field theories of strong interactions%
\footnote{
In the sense first suggested by Weinberg
\cite{WeinbEff};
see
\cite[Sec.~1]{POMI}
and
\cite[Sec.~2]{AVVV2}
for comments concerning the quantization scheme.
}.
The reason to work with Dyson's scheme is that this is the only known
way that allows one to combine Lorentz invariance, cluster
decomposition principle, and unitarity with general postulates of
quantum mechanics
\cite{WeinMONO}.
Thus, the problems we have to do with include the infinite number of
graphs to be summed at each loop (including the tree) level and the
problem of the required number of renormalization conditions, since in
such a theory one needs to fix an infinite number of parameters to be
able to calculate amplitudes. In
\cite{VVV,AVVV1}
it has been shown that already the requirement of summability of tree
graphs --- the tree level amplitude --- leads to strong restrictions
on the coupling constants of a theory. However, in those articles many
of theoretical statements, like the
{\em meromorphy} and
{\em polynomial boundedness}
of the tree amplitude, were taken as postulates and only some general
arguments in their favor were given. With these assumptions it turned
out possible to obtain
{\em bootstrap equations}
for masses and coupling constants of
$\pi\pi$ and
$\pi K$
resonances in nice agreement with the experimental data. The main tool
used to derive those equations is the
{\em Cauchy expansion},
based on the celebrated Cauchy integral formula, which allows one to
represent the tree level amplitude as well defined series in a given
domain of the space of kinematical variables
\cite{VVV-QFTHEP,AVVV1}.

In subsequent publications we fill some gaps left in the previous
analysis as well as discuss new concepts. Thus, in
\cite{POMI,MENU}
we suggested the notion of
{\em minimal parametrization}
and in
\cite{AVVV2}
the relevant reduction theorem is proven. This explains why it is
sufficient to consider only the minimal
(``quasi-on-shell'')
vertices at each loop order of the perturbation theory --- the fact
implicitly used in
\cite{VVV,AVVV1}
to parametrize amplitudes. Besides, in
\cite{POMI}
we discuss what we call the
{\em localizability principle} ---
the philosophy which, in particular, serves as the background for
meromorphy and polynomial boundedness of the tree level amplitude.
However, the last point is still not explained clear enough. A brief
discussion can be found in
\cite{HSQCD}
where we introduce so-called
{\em summability} and
{\em asymptotic uniformity}
principles, and we plan to treat it in more detail in a separate
publication. Here we do not focus on this and just assume it is known
that the tree level
$\pi N \to \pi N$
amplitude is a polynomially bounded meromorphic function of any pair
of independent kinematical variables.

This talk is mainly devoted to the derivation of the bootstrap
equations for
$\pi N$
elastic scattering and the results of comparison with the experimental
data.

\section{Structure of the amplitude and minimal vertices}
\label{tensor}

The amplitude
$M_{a \alpha}^{b \beta}$ of the reaction
\begin{equation}
\pi_{a} (k) + N_{\alpha} (p, \lambda) \to
\pi_{b} (k') + N_{\beta} (p', \lambda')
\label{2.1}
\end{equation}
can be presented in the following form:
\begin{equation}
M_{a \alpha}^{b \beta} = i{(2\pi)}^4 \delta (k+p-k'-p')
\left\{
\delta_{ba}\delta_{\beta\alpha} M^+ +
i\varepsilon_{bac}(\sigma_{c})_{\beta\alpha} M^{-}
\right\} \;\; .
\label{2.2}
\end{equation}
Here
\[
M^{\pm} = \overline{u}(p',\lambda') \left\{
A^{\pm}+\left(\frac{\s{k}+\s{k'}}{2}\right) B^{\pm}
\right\} u(p,\lambda)\;\; ,
\]
$a, b = 1,2,3$
and
$\alpha,\beta = 1,2$
stand for the isospin indices,
$\lambda, \lambda'$ ---
for polarizations of the initial and final nucleon, respectively,
$\overline{u}(p',\lambda')$ and $u(p,\lambda)$ ---
for Dirac spinors and
$\sigma_c,\; c=1,2,3$ ---
for Pauli matrices. The invariant amplitudes
$A^{\pm}$
and
$B^{\pm}$
are the functions of an arbitrary pair of Mandelstam variables
$s\equiv (p+k)^2$, $t\equiv (k-k')^2$,
and
$u\equiv (p-k')^2$.

\begin{figure}[ht]
\begin{center}
\begin{picture}(350,20)(25,-10)
\put(0,0){
\begin{picture}(100,20)(0,0) 
\put(0,-5){\shortstack{$\displaystyle\sum_{ \rm vertices \atop
\mbox{} }^{\infty}$}}
\put(40,0){\circle*{3}}
\put(40,0){\line(-1,1){10}}
\put(40,0){\line(-1,-1){10}}
\put(30,-10){\vector(1,1){7}} 
\put(40,0){\line(1,-1){10}}
\put(40,0){\vector(1,-1){7}} 
\put(40,0){\line(1,1){10}}
\put(60,-10){\shortstack{\boldmath$,$}}
\end{picture}
}

\put(100,0){
\begin{picture}(100,20)(0,0) 
\put(0,-5){\shortstack{$\displaystyle\sum_{ \rm vertices, \atop
\rm resonances }^{\infty}$}}
\put(40,0){\circle*{3}}
\put(40,0){\line(-1,1){10}}
\put(40,0){\line(-1,-1){10}}
\put(30,-10){\vector(1,1){7}} 
\multiput(40,0)(1,0){20}{\circle*{2}}  
\put(40,0.5){\vector(1,0){13}} 
\put(40,-0.5){\vector(1,0){13}} 
\put(45,-13){\shortstack{$R_s$}}
\put(60,0){\circle*{3}}
\put(60,0){\line(1,1){10}}
\put(60,0){\line(1,-1){10}}
\put(60,0){\vector(1,-1){7}} 
\put(80,-10){\shortstack{\boldmath$,$}}
\end{picture}
}

\put(200,0){
\begin{picture}(100,20)(0,0) 
\put(0,-5){\shortstack{$\displaystyle\sum_{ \rm vertices, \atop
\rm resonances}^{\infty}$}}
\put(50,-10){\circle*{3}}
\put(50,-10){\line(-1,-1){10}}
\put(40,-20){\vector(1,1){7}} 
\put(50,-10){\line(1,-1){10}}
\put(50,-10){\vector(1,-1){7}} 
\multiput(50,-10)(0,1){20}{\circle*{2}}  
\put(53,-5){\shortstack{$R_t$}}
\put(50,10){\circle*{3}}
\put(50,10){\line(-1,1){10}}
\put(50,10){\line(1,1){10}}
\put(75,-10){\shortstack{\boldmath$,$}}
\end{picture}
}

\put(300,0){
\begin{picture}(100,20)(0,0)  
\put(0,-5){\shortstack{$\displaystyle\sum_{ \rm vertices, \atop
\rm resonances}^{\infty}$}}
\put(40,0){\circle*{3}}
\multiput(40,0)(1,0){20}{\circle*{2}}  
\put(40,0.5){\vector(1,0){13}} 
\put(40,-0.5){\vector(1,0){13}} 
\put(45,-13){\shortstack{$R_u$}}
\put(60,0){\line(1,-1){10}}
\put(60,0){\vector(1,-1){7}} 
\put(60,0){\circle*{3}}
\put(30,-10){\line(1,1){15}}
\put(30,-10){\vector(1,1){7}} 
\put(55,15){\line(1,1){10}}
\put(60,0){\line(-1,1){25}}
\put(55,5){\oval(20,20)[tl]}
\put(80,-10){\shortstack{\boldmath$.$}}
\end{picture} }

\end{picture}
\end{center}
\caption{Tree level graphs:
$R_s$, $R_t$ and $R_u$ stand for all admissible resonances in
$s$-, $t$-, and $u$-channels, respectively; summation over all
possible kinds of vertices is implied, though the summation order is
still unspecified.
\label{1f}}
\end{figure}
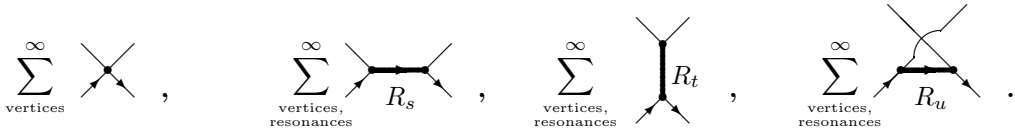

Since we rely upon the Dyson's scheme, we must construct the amplitude
order by order, starting from the tree level. To construct the tree
level amplitude, one needs to collect contributions of the graphs
shown in
Fig.~\ref{1f}.

We deal with the effective theory. This means that the interaction
Hamiltonian%
\footnote{
In
\cite{AVVV2}
it is explained why it is preferably to use the effective interaction
Hamiltonian (which, of course, can be written in a Lorentz covariant
form), rather than Lagrangian when constructing a theory with
unlimited number of field derivatives.
},
(in the interaction picture) contains
{\em all}
the terms consistent with (algebraic) symmetry properties of strong
interactions; there are no limitations on the number and the order of
field derivatives. Besides, to avoid model dependence we reserve the
possibility to work with infinite number of resonance fields.
Altogether this means that the number of items contributing to the
tree level amplitude is actually infinite. This creates a problem:
there is no guarantee that the sum of these terms taken
``as it is''
converges, at least, in some domain of the space of kinematical
variables. Actually, as it usually happens with functional series, it
{\em diverges}
in general, and the tree level amplitude simply
{\em does not exist}
until one attracts a guiding principle to fix the order of summation.

The way out of this difficulty was pointed out in
\cite{VVV}--\cite{AVVV2}.
It consists of switching to the
{\em minimal}\,
parametrization and using the method of Cauchy forms in various
regions of the complex space of kinematical variables. The important
advantage of this approach is that it results in uniformly converging
series of singular terms defining the tree level amplitude as the
polynomially bounded meromorphic function --- no kind of singularities
but simple poles can appear on this way. To construct the Cauchy form
for the tree level amplitude, one only needs to fix the residues at
corresponding poles and properly choose the bounding polynomial degree
--- it happens quite sufficient for fixing the amplitude up to few
unknown functions. These latter functions, in turn, can usually be
found from the
{\em bootstrap equations}.

By the very definition, the residues at poles of the tree level
amplitude are nothing but the relevant spin sums (numerators of
minimal propagators) dotted by the minimal triple coupling constants
(those at minimal triple vertices). This is so just because only the
minimal vertices survive on the mass shell and, therefore, do not
cancel propagator's numerator. In
\cite{VVV-QFTHEP}
it is shown that contribution of vertices with four (and more)
external lines does not require fixing by independent renormalization
prescriptions. That is why in our case the pole graph parameters ---
minimal triple vertices and masses, together with asymptotic regime
and self-consistency requirements completely determine the tree level
amplitude and, as a by-product, the contribution of the contact
four-leg vertices.

The Hamiltonian density monomials corresponding to the minimal triple
vertices contributing to the amplitude of elastic pion-nucleon
scattering are listed below%
\footnote{
See
\cite{AVVV2,HSQCD}
for rigorous definition of the minimal vertex and minimal
parametrization.
}.
Note, that there is only a
{\em finite}
number of minimal triple vertices for each resonance field with given
quantum numbers. In this article we consider a concrete process and
employ experimental data, therefore it is natural to input some
phenomenology already at this stage. Namely, we imply that there are
no resonances with isospin
$I \geq 2$.
Later, when coming to the amplitude calculation, we make another
assumption --- that the number of resonances with the same mass is
finite. These restrictions are kept automatically when one substitute
phenomenological data. On the other hand, they do not affect the
mathematics and can easily be relaxed.

Henceforth we use the Rarita-Schwinger formalism for the higher spin
fields. In the formulae below
$J$
denotes spin and
$I$ ---
the isotopic spin (we drop the isospin indices and write isovectors in
boldface to form scalar and vector products),
$P$
is the parity, and
$\cal N$
is the
``normality''
of the corresponding resonance, the latter equals
$(-1)^{(J-1/2)} P$
for baryons and
$(-1)^{J} P$
for mesons. Besides,
$\pi$
is the pion field and
$N$
is the nucleon field. The baryon resonance fields are denoted by
$\widehat{R}$
and
$R$ ($I = 1/2$), $\widehat{\Delta}$
and
$\Delta$ ($I = 3/2$)%
\footnote{
Note that the experimental
$\Delta(1.23)$
resonance
($I,J = 3/2$, $P = +1$)
has
${\cal N} = -1$
and thus belongs to the type
$\widehat{\Delta}$
in our notations.
},
the mesons are
$S$ ($I = 0$),
and
$V$ ($I = 1$).
In the minimal parametrization the vertices (Hamiltonian monomials) we
need are the following:

\noindent
\begin{description}
\item{$J=l+{1\over 2}\; (l=0,1,2\ldots) $}
\begin{description}
\item{$I={1\over 2},\; {\cal N}=-1$: \quad}
$
g_{\widehat{R}}
\overline{N}\bo{\sigma}
\widehat{R}_{\mu_1\ldots\mu_l}
\partial^{\mu_1}\!\!\!\!\ldots\partial^{\mu_l}\bo{\pi}
+ H.c.
$
\item{$I={1\over 2},\; {\cal N}=+1$: \quad}
$
ig_{R}
\overline{N} \bo{\sigma} \gamma_5
R_{\mu_1\ldots\mu_l}
\partial^{\mu_1}\!\!\!\!\ldots\partial^{\mu_l}\bo{\pi}
+ H.c.
$
\item{$I={3\over 2},\; {\cal N}=-1$: \quad}
$
g_{\widehat{\Delta}}
\overline{N} P_{3\over 2}
\widehat{\Delta}_{\mu_1\ldots\mu_l}
\partial^{\mu_1}\!\!\!\!\ldots\partial^{\mu_l} \bo{\pi}
+ H.c.
$
\item{$I={3\over 2},\; {\cal N}=+1$: \quad}
$
ig_{\Delta}
\overline{N} \gamma_5 P_{3\over 2}
\Delta_{\mu_1\ldots\mu_l}
\partial^{\mu_1}\!\!\!\!\ldots\partial^{\mu_l} \bo{\pi}
+ H.c.
$
\end{description}
\item{$J=0,2,\ldots,\; I=0,\; P=+1$:}
\begin{description}
\item
\begin{center}
$
{1\over 2}g_{S\pi\pi}
S_{\mu_1\ldots\mu_J}
(\bo{\pi}\cdot\partial^{\mu_1}\!\!\!\!\ldots\partial^{\mu_J}\bo{\pi})
\; ,
$
\end{center}
\item
\begin{center}
$
\left[
g^{(1)}_{NNS}
\overline{N}\partial_{\mu_1}\!\!\ldots\partial_{\mu_J} N
+ ig^{(2)}_{NNS} J
\partial_{\mu_1}\!\!\ldots\partial_{\mu_{J-1}}
\overline{N}\gamma_{\mu_J} N
\right]
S^{\mu_1\ldots\mu_J}\; ,
$
\end{center}
\end{description}
\item{$J=1,3,\ldots,\; I=1,\; P=-1$:}
\begin{description}
\item
\begin{center}
$
{1\over 2}g_{V\pi\pi}
\bo{V}_{\mu_1\ldots\mu_J}
(\bo{\pi}\times\partial^{\mu_1}\!\!\!\!\ldots\partial^{\mu_J}\bo{\pi})
\; ,
$
\end{center}
\item
\begin{equation}
\left[
i g^{(1)}_{NNV}
\overline{N} \bo{\sigma}
\partial_{\mu_1}\!\!\ldots\partial_{\mu_J} N
+ g^{(2)}_{NNV} J \overline{N} \gamma_{\mu_J}
\bo{\sigma} \partial_{\mu_1}\!\!\ldots\partial_{\mu_{J-1}} N
\right]
\bo{V}^{\mu_1\ldots\mu_J}\; .
\label{hamiltonian}
\end{equation}
\end{description}
\end{description}
Here
$g^{\cdots}_{\ldots}\equiv g^{\cdots}_{\ldots} (J,I,P)$
stand for real coupling constants, and
$\;\sigma_c\; (c=1,2,3)$ ---
for Pauli matrices.
\[
P_{3\over 2} \equiv
\left(P_{3\over 2}\right)_{a \alpha b \beta} =
{2 \over 3}\left\{
\delta_{\alpha\beta}\delta_{ab} -
{i\over 2} \varepsilon_{abc}
\left( \sigma_c \right)_{\alpha\beta}
\right\},\;
(a, b = 1,2,3,\; \alpha,\beta=1,2)
\]
is the projecting operator on the states with isospin
$I = 3/2$,
$a,\alpha,b,\beta$
being the isotopic indices.

There are no more minimal triple vertices contributing to the tree
level amplitude. In
\cite{HSQCD}
it is explained that the
$g$'s
above appears to be not just minimal, but also the
{\em resultant}
tree level couplings. As pointed out in
\cite{AVVV2},
the resultant parameters are the natural candidates to impose the
renormalization prescriptions on, under the condition that the
renormalization point is taken on shell and the
{\em renormalized perturbation theory}
is used. In this scheme the action is written in terms of
{\em physical}
parameters plus counterterms, the latter ones being tuned in such a
way that the values of physical parameters remain unchanged after the
renormalization. So, we imply that the Feynman rules are written in
the form of physical part plus counterterms at every loop order and it
is the
{\em real parts of physical masses}
that appear in bare propagators. Simply speaking, we impose the
following set of RP's:
\[
\bo{\rm Re}\ V(p_1,p_2,p_3) = G_{phys} {\rm \; at \;}
p_i^2 = M_{i_{phys}}^2,
\]
and
\[
\bo{\rm Re}\ \Sigma(p) = 0 {\rm \; at \; }
p^2 = M_{phys}^2,
\]
for every self-energy
$\Sigma$
and every three-leg vertex
$V$.
Now we are at tree level, thus there are no counterterms relevant,
therefore the couplings
$g$
are also physical --- (in principle) measurable, or
{\em renormalized} ones.

There is no phenomenological evidence that the mass spectrum and spin
values of resonances are bounded from above. Therefore we need to
reserve the possibility to work with an infinite set of resonances of
arbitrary high spin value. In other words, there is still infinite
number of minimal couplings coming even from three-leg vertices. One
of the main points of our work is that these couplings, or, the same,
renormalization conditions fixing them, are
{\em not independent}:
there are
{\em self-consistency conditions}
that restrict their values. Thus, to make the perturbation theory
self-consistent, one should make sure that all the restrictions are
met at any loop order. The
{\em bootstrap}
constraints discussed below form a part of such restrictions.

\section{Cauchy forms}
\label{Cauchy}

\begin{figure}[ht]
\begin{center}
\begin{picture}(220,220)(-100,-70)


\thicklines

\put(15,110){\vector(-1,-2){69}}       
\put(-60,0){\vector(1,0){140}}         
\put(55,-30){\vector(-1,2){76}}        
\put(70,-10){\shortstack{$\bo{\nu_t}$}}     
\put(-30,107){\shortstack{$\bo{\nu_u}$}}    
\put(-50,-30){\shortstack{$\bo{\nu_s}$}}    

\put(0,0){\vector(0,1){15}}         
\put(2,13){\shortstack{$\bo{t}$}}   

\put(-20,40){\vector(2,-1){12}}     
\put(-13,27){\shortstack{$\bo{s}$}} 

\put(20,40){\vector(-2,-1){12}}     
\put(0,35){\shortstack{$\bo{u}$}}   

\thinlines

\put(0,0){\circle*{2}}              
\put(20,40){\circle*{2}}            
\put(-20,40){\circle*{2}}           


\multiput(-40,-20)(1,2){60}{\circle*{0.2}}  
\multiput(-60,10)(2,0){60}{\circle*{0.2}}   
\multiput(40,-20)(-1,2){60}{\circle*{0.2}}  

\multiput(-60,-20)(1,2){70}{\circle*{0.2}}  
\multiput(-60,-10)(2,0){60}{\circle*{0.2}}  
\multiput(60,-20)(-1,2){70}{\circle*{0.2}}  

\multiput(-55,-10)(10,20){2}{\line(1,0){20}}  
\multiput(-55,-10)(1.5,0){14}{\line(1,2){10}} 

\multiput(35,-10)(-10,20){2}{\line(1,0){20}}  
\multiput(35,-10)(1.5,0){14}{\line(-1,2){10}} 

\multiput(0,60)(-0.7,1.4){15}{\line(1,2){10}} 
\multiput(0,60)(10,20){2}{\line(-1,2){10}}    

\put(-54,55){\shortstack{$B_s$}}  
\put(-27,37){\line(-1,1){15}}     

\put(-5,-30){\shortstack{$B_t$}}  
\put(-5,-5){\line(1,-3){5}}       

\put(41,53){\shortstack{$B_u$}}   
\put(27,37){\line(1,1){15}}       

\put(45,5){\line(1,3){6}}         
\put(47,27){\shortstack{$D_s$}}   

\put(5,80){\line(3,1){20}}        
\put(26,85){\shortstack{$D_t$}}   

\put(-45,5){\line(-1,3){7}}      
\put(-61,29){\shortstack{$D_u$}}  


\multiput(50,40)(18,-9){3}{
\begin{picture}(40,40)(8,0)
\multiput(0,0)(-7,-14){6}{\line(-1,-2){5}} 
\end{picture}
}
\multiput(70,-30)(6,-3){4}{\circle*{1}}
\put(45,-65){\shortstack[l]{$s=M_s^2$ \tiny poles }}

\multiput(-80,75)(0,20){3}{
\begin{picture}(40,40)(0,0)
\multiput(0,0)(15,0){10}{\line(1,0){10}} 
\end{picture}
}
\multiput(2,130)(0,5){3}{\circle*{1}}  
\put(15,130){\shortstack[l]{$t=M_t^2$ \tiny poles }}

\multiput(-50,40)(-18,-9){3}{
\begin{picture}(40,40)(0,0)
\multiput(0,0)(7,-14){6}{\line(1,-2){5}} 
\end{picture}
}
\multiput(-70,-30)(-6,-3){4}{\circle*{1}}
\put(-98,-65){\shortstack[l]{$u=M_u^2$ \tiny poles }}

\end{picture}
\end{center}
\caption{Mandelstam plane: three different Cauchy series
uniformly converge in three different bands
$B_s$, $B_t$ and
$B_u$ (bounded by dotted lines). The domains
$D_s$, $D_t$, $D_u$
of the band intersections are hatched. The dashed lines show the pole
positions in the relevant variables.
\label{fig:2} }
\end{figure}
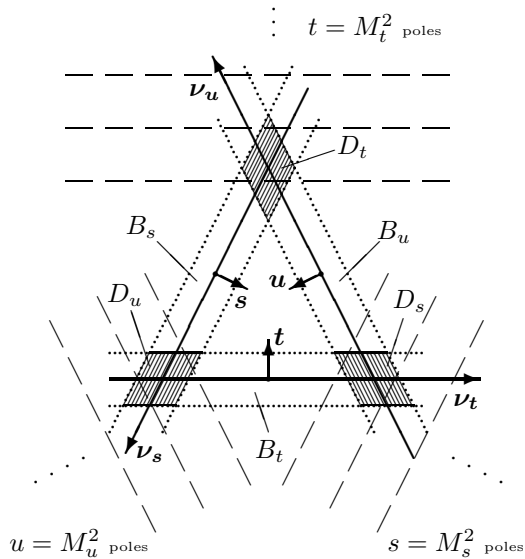

In this Section we show how to construct the Cauchy forms
(series): the well-defined expressions for the tree level amplitudes
$A^{\pm}$
and
$B^{\pm}$
in three mutually intersecting bands
($B_s$, $B_t$
and
$B_u$)
of the Mandelstam plane
(Fig.~\ref{fig:2}).
\noindent
In addition to conventional Mandelstam variables
$s,t,u$
it is convenient to define
$\nu_s \equiv u-t$, $\nu_t \equiv s-u$
and
$\nu_u \equiv t-s$.
Every pair
$(x, \nu_x)$
presents the orthogonal coordinate system in the band
(layer)
\[
B_x\{x \in {\bf R},\,  x \sim 0;\,  {\nu}_x \in {\bf C}\}, \;
(x=s,t,u),
\]
where the corresponding series converges. The Cauchy form arise as a
result of the application of the Cauchy integral formula to the tree
level amplitude in one of those layers.

Let us, for definiteness, consider the band
$B_t$.
Referring to the talk
\cite{VVV-QFTHEP}
and keeping in mind that, according to the summability principle
\cite{HSQCD},
the only singularities of the tree level amplitude are the poles
appearing in the tree-level exchange graphs, we schematically
rewrite
\cite[Eq.~2]{VVV-QFTHEP}
as
\begin{equation}
\left.
X (\nu_t, t)
\right|_{B_t}
= \sum \limits_{k=0}^{N}
\frac{1}{k!} \alpha_k (t) \nu_t^k
+
\sum\limits_n^\infty
\left\{\frac{p_n (t)}{s - M_{u_n}^2} + \frac{q_n (t)}{u - M_{s_n}^2} -
\sum \limits_{m=0}^{N}\beta_{n,m}(t) \nu_t^k
\right\},
\label{Cauchy-form}
\end{equation}
where
$X$
stands for any one of the invariant amplitudes
$A^{\pm}$
or
$B^{\pm}$
and the summation in order of increasing mass is implied. We call the
above expression as
{\em Cauchy form}.
In
\cite{VVV-QFTHEP}
it is shown that, as long as the bounding polynomial degree
$N$
is given, the terms in brackets: poles and the coefficients
$\beta$
of the so-called
{\em correcting polynomials}
are easily expressed via the minimal triple couplings
(\ref{hamiltonian})
and masses
$M_{s_n}$
and
$M_{u_n}$
of the
$s$- and
$u$-channel
resonances. The
$\alpha$'s
are still unknown at this stage. Analogous forms can be easily written
for the tree level amplitude in the cross-symmetric bands
$B_s$
and
$B_u$.

According to the
{\em asymptotic uniformity}
principle
\cite{HSQCD},
one should admit the tree level amplitude asymptotic behavior to be of
the same order as that dictated be the corresponding Regge intercept.
One can check that in our case all the invariant
amplitudes has decreasing asymptotics in the bands
$B_s$
and
$B_u$
(behave like negative powers of
$\nu_s$
and
$\nu_u$,
respectively), while in
$B_t$: $A^{+}\sim\nu_t^{1}$, $A^{-}\sim\nu_t^{0.5}$,
$B^{+}\sim\nu_t^{0}$,
and
$B^{-}$ again behave like negative power of
$\nu_t$.
Attracting the odd-even properties, we conclude that only the form for
$A^{+}$
in the band
$B_t$
requires correcting polynomials and external part%
\footnote{
According to the asymptotics,
$A^{-}$ in
$B_t$ could also require the correcting polynomials and external part
of zeroth order in
$\nu_t$.
However, the spectrum is symmetric with respect to
$\nu_t \rightarrow -\nu_t$
transformation and
$A^{-}$
is odd in
$\nu_t$.
Thus, the zeroth order polynomials in the corresponding Cauchy form
are forbidden (cancel in pairs). Similarly,
$A^{+}$
is even in
$\nu_t$,
hence in
$B_t$
its correcting polynomials degree equals
$0$ ---
not
$1$
as one could expect from the value of relevant intercept.
}
$\alpha(t)$.
Cauchy series for all the other amplitudes are just the sums of poles
appearing in the corresponding channels (layers).

We have no space here to specify the exact amplitude expansion in
each given band, and later on just briefly sketch the way one obtains
the bootstrap equations. However, before discussing the bootstrap, it
is useful to summarize what we have in hands up to now.

The Cauchy forms above are written for the tree level amplitudes. They
are the series of special type, each one being valid (convergent) in
certain layer on the Mandelstam plane. The only unknown component is
$\alpha (t)$
appearing in
$A^+$
expression in
$B_t$,
all the other items are known functions of kinematical variables,
triple couplings and masses. In each separately taken form the pole
parts originates from those exchange graphs, that contain the poles
crossing the band in which the form is valid. Neither cross channel
poles nor contact graph (four-leg vertices) contributions appear
explicitly%
\footnote{
Except the trivial cases, each graph with resonance exchange contains
a polynomial contribution, as well as the pole. Our minimal (on-shell)
parametrization allows one to get rid of the polynomial part, but, of
course, the contact vertices still contribute.
}.
We can formulate it stronger: in almost all of the amplitude
expressions in the given channel (layer) the cross channel exchange
graphs and four-point vertices do not show up at all --- the
amplitude is written as just the sum of poles in the given layer. For
example, there are no
$t=M^2$
poles in
(\ref{Cauchy-form}).
It shows that in many cases the asymptotic regime requires the
cancellation of contributions coming from cross channel poles with
that following from the contact graphs. What is necessary to stress is
that it is not always the case, and
$A^+$
expansion in
$B_t$
gives an excellent illustration: the
$\alpha$
and the correcting polynomials in the Cauchy form for
$A^+$
can be regarded as a joint effect of contact and cross channel
exchange graphs.

\section{Bootstrap equations and experimental data}

As it is seen from the
Fig.~\ref{fig:2},
each pair of the bands
$B_s$, $B_t$
and
$B_u$
has non-empty intersection domain:
$D_s$, $D_t$
and
$D_u$.
To obtain the bootstrap constraints we just equate each two Cauchy
series for a given amplitude in the domain where both forms are valid%
\footnote{
Namely
$D_s$, $D_t$,
or
$D_u$.
However, in the case under consideration the equations one obtains in
$D_u$
are equivalent to those from
$D_s$
due to the symmetry of the process. Note also, that these equations
are nothing but the requirements of crossing symmetry.
}.
After this is done, the function
$\alpha(t)$
mentioned in the previous Section can be expressed in terms of the
parameters of cross channel Cauchy form (minimal triple couplings and
masses). It is at this stage that the tree level amplitude turns out
to be completely specified.

Expanding the resulting (bootstrap) equations in powers of kinematical
variables (for example, around the
$t=u=0$
point for the equalities valid in
$D_s$),
one obtains an infinite set of numerical equations for the resultant
couplings and masses with no kinematics involved anymore.

The bootstrap equations have rather complicated form, but what is
necessary to mention again, is that according to the consideration in
the end of
Sec.~\ref{tensor},
all the coupling constants and masses that enter are the physical
(renormalized) ones. This means that one can check the resulting
relations using the experimental values of the known resonances
parameters.

Before turning to numerical analysis it looks appropriate to discuss
here the meaning of the terms
``mass''
and
``width''
often used to specify the parameters describing a resonance.

In our approach we are only allowed to calculate the amplitude in
question precisely at a given loop order; no kind of partial
resummation is allowed%
\footnote{
The latter can lead, e.g., to the violation of perturbative unitarity.
}.
In particular, we are not allowed to perform Dyson's resummation for
the propagator. This means that the theoretical expression for the
amplitude of, say, elastic pion-nucleon scattering should be
constructed from the propagators corresponding to simple poles at real
values of the momentum squared. This circumstance creates a problem
when one needs to compare the (finite loop order) theoretical
expression with experimental data. To circumvent this difficulty one
needs to exclude from a consideration those data which correspond to
small vicinities of the resonance poles. Technically, this is just a
problem of appropriate organization of the fitting procedure; it is
easily solvable. In contrast, from the purely theoretical viewpoint
this is a problem of definition of the parameters describing a
resonance. We would like to stress that the notion of resonance only
acquires meaning in the framework of a particular perturbation theory.
If we could construct the complete non-perturbative expressions for
S-matrix elements we would never need to use the term ``resonance''.

Since the partial resummation is not allowed, one cannot construct the
finite order amplitude with complex poles. Hence, the customary
Breit-Wigner description of resonances (mass and width as real and
imaginary parts of the pole position, respectively) looses its meaning
in terms of the resultant parameters%
\footnote{Generally speaking, this is just a consequence of the fact
that the notion of resonance is ill-defined in the framework of
perturbative quantum field theory
\cite{WeinMONO}.
}.
Instead, we use the term
``physical mass''
to denote the parameter
$M_i$
appearing in the denominator of the free particle propagator (recall
that we rely upon the scheme of renormalized perturbation theory).
This implies using the renormalization prescription given in
Sec.\ref{tensor}

When fitting data with the finite loop order amplitude one obtains the
values of relevant physical couplings which, in turn, can be used to
compute (formally!) the decay amplitude and, hence, the resonance
width. Thus, in principle, it is possible to avoid using the
ill-defined term
``width''
and work in terms of coupling constants instead. However, when
comparing our bootstrap equations with experiment, we are forced to
use the numbers quoted, say, in
\cite{PDG}.
Those numbers are given in terms of mass and width corresponding to
the parameters describing the T-matrix poles at an unphysical sheet of
the complex energy plane. In many other sources the data on resonances
are quoted in terms of Breit-Wigner or K-matrix parameters. Clearly,
in the case of well-separated narrow resonance the different methods
of parametrization of the amplitude (including that based on field
theory models operating directly with coupling constants) result in
approximately the same values of mass and coupling constant. This is,
however, not true with respect to broad resonances like famous light
scalars%
\footnote{An excellent discussion of this point can be found in the
series of papers
\cite{Schechter}.
}.
In this latter case the connection between the mass parameters
appearing in propagators, coupling constants and the parameters
describing poles in complex energy plane (real and imaginary part)
becomes very complicated. In other words, the values of mass and
coupling constant extracted from the same data set with the help of
different theoretical amplitudes may differ considerably. Surely, this
is explained by well known difficulties of analytic continuation of
the approximate expressions obtained in the framework of certain
perturbation scheme. This is the reason why we find it necessary to
clarify the terminology.

As an example of comparison with experimental data we shall mention a
calculation of the ratio for two
$\rho$-meson
coupling constants. The quantities
$G^T_{NN\rho}$
and
$G^V_{NN\rho}$
were defined and fitted in
\cite{Nagels}
as couplings in the effective Hamiltonian
\begin{equation}
H^{NN\rho}_{\rm eff} = -\overline{N}
\left[
G^V_{NN\rho}\gamma_\mu\bo{\rho}^\mu -
G^T_{NN\rho}{\sigma_{\mu\nu}\over 4 m}
\left(\partial^\mu\bo{\rho}^\nu -\partial^\nu\bo{\rho}^\mu\right)
\right]
{1\over 2} \bo{\sigma} N\; ,
\end{equation}
where
$\sigma_a $
are Pauli matrices and
$m$
is the proton mass. The existing experimental data
\cite{Nagels}
give:
\begin{equation}
{ G^T_{NN\rho} \over G^V_{NN\rho}} \approx 6.1\ ,\;\;
\frac{G_{\pi \pi \rho}G^V_{NN\rho}}{4\pi} \approx 2.4\ , \;\;
G_{\pi \pi \rho} \approx 6.0\ .
\label{gt/gvexp}
\end{equation}
These values can be deduced from the bootstrap equations for
$\pi N$
elastic scattering with
$15\%$
accuracy%
\footnote{
The problem of the convergence rapidity of various types of the
bootstrap equations is discussed in
\cite{KSTSH-QFTHEP}.
}.
It should be also mentioned here that the same value for the first
ratio was obtained during the similar analysis of
$KN$
elastic scattering
\cite{HSQCD}.

Besides, we checked the tree level bootstrap equations for
$\pi \pi$
and
$\pi K$
elastic scattering amplitudes (see
\cite{AVVV1}
and references therein). Similar calculations were performed for the
cases of
$\pi N$
\cite{MENU}
and, recently,
$KN$
elastic scattering
\cite{HSQCD}.
There were no contradiction found so far, and in most cases examined
the experimental data seem to support our approach nicely.

Apart from the question of formal compatibility with experiment, there
is a question of efficiency. One can ask how many loops should be
taken into account and how many parameters fixed to obtain the
amplitude that could fit well the data at least in some kinematical
region. To check this point we performed a calculation of low energy
coefficients%
\footnote{
Taylor expansion coefficients around the crossing symmetry point.
}
for the
$\pi N$
amplitude. This coefficients measured and fitted in
\cite{Nagels}
are reproduced in our approach with very good accuracy already at tree
level%
\footnote{
Of course, it is partly because this region is relatively far from the
branch points. In case if the latter appears close to the investigated
region one should necessary include loops.
},
and to gain reasonable precision it is enough to specify the
parameters of just few lightest resonances. The results of this
analysis were summarized in
\cite{MENU};
the details will be published elsewhere.

I am grateful to K.~Semenov-Tian-Shansky and V.~Vereshagin for helpful
collaboration as well as to V.~Cheianov, H.~Nielsen, S.~Paston,
J.~Schechter, A.~Vasiliev and M.~Vyazovsky for stimulating
discussions. This work is supported by INTAS (project 587, 2000),
Ministry of Education of Russia (Programme ``Universities of Russia'')
and L.~Meltzers H\o yskolefond (Studentprosjektstipend, 2004).


\end{document}